\documentclass[epj,nopacs]{svjour}
\usepackage{epsfig}
\usepackage{amsfonts}
\usepackage{amsmath}
\usepackage{bbm}
\usepackage{cite}
\allowdisplaybreaks[1]

\input pix.sty

\renewcommand{\eq}{eq.~}
\renewcommand{\eqs}{eqs.~}

\newcommand{\rmO}{{\mathcal{O}}}

\def\lsi{\raise0.3ex\hbox{$<$\kern-0.75em\raise-1.1ex\hbox{$\sim$}}}
\def\gsi{\raise0.3ex\hbox{$>$\kern-0.75em\raise-1.1ex\hbox{$\sim$}}}

\newcommand{\gsim}{\mathop{\gsi}}

\newcommand{\nB}{n_\rmii{B}}

\newcommand{\rmii}[1]{{\mbox{\tiny\rm{#1}}}}

\newcommand{\im}{\mathop{\mbox{Im}}}

\newcommand{\Tint}[1]{{\hbox{$\sum$}\!\!\!\!\!\!\!\int\,}_{\!\!\!\!\raise-0.9ex\hbox{$\scriptstyle{#1}$}}}
\newcommand{\Tinti}[1]{{{\Sigma}\!\!\!\!\raise0.3ex\hbox{$\int$}_\rmii{${#1}$}}}

\newcommand{\bi}{\begin{itemize}}
\newcommand{\ei}{\end{itemize}}
\newcommand{\hide}[1]{ }

\newcommand{\deltabar}{\delta\!\!\!\raise0.7ex\hbox{--}\,}
\makeatletter \@addtoreset{equation}{section} \makeatother
\renewcommand{\theequation}{\arabic{section}.\arabic{equation}}

\begin{document}

\flushbottom
\sloppy

%
%
%
\title{
  Hydrodynamic fluctuations from a weakly coupled scalar field 
}

\author{G.~Jackson \and M.~Laine} 
\institute{
AEC, Institute for Theoretical Physics, 
University of Bern,  
Sidlerstrasse 5, CH-3012 Bern, Switzerland} 

\date{March 2018}

%
%
%
%
%
\abstract{%
Studies of non-equilibrium dynamics of first-order
cosmological phase transitions may involve a scalar field 
interacting weakly with the energy-momentum tensor of a
thermal plasma. At late times, when the scalar field is approaching
equilibrium, it experiences both damping and thermal fluctuations.  
We show that thermal fluctuations induce a shear viscosity and a
gravitational wave production rate, and propose that including this
tunable contribution may help in calibrating the measurement of the
gravitational wave production rate in hydrodynamic simulations. 
Furthermore it may enrich their physical scope, permitting in particular 
for a study of the instability of growing bubbles.
}

%
 
%
%

\maketitle

%
\section{Introduction}

With the planning of the LISA interferometer under way,
it has become timely to consider gravitational
wave production from cosmological phase 
transitions~\cite{lisa_transition}.
This process is dominated by 
non-equilibrium sources, with a considerable
contribution originating from a late stage with sound waves
and/or turbulence (cf.\ ref.~\cite{weird} for an overview
of recent work).
Eventually this motion terminates and the system
reaches thermal equilibrium. In the equilibrium state
the production of gravitational waves
continues through thermal fluctuations~\cite{jacopo} 
but the magnitude of this component
is in general much below the observable level. 

A phase transition proceeds through the nucleation
and subsequent growth, collisions, and coalescence 
of bubbles of the low-temperature phase. As they are
growing, the bubble walls reach a steady velocity, 
because of friction~(cf.\ ref.~\cite{db_gdm} and references therein).
Friction can be represented by 
a dissipative coefficient in the equation of motion
for the order parameter of the transition
(``scalar field''). 
The fluctuation-dissipation
theorem asserts that dissipation 
implies the presence of fluctuations. The purpose
of this study is to show how thermal fluctuations
of the scalar field can be included
in a framework frequently used for numerical simulations~\cite{weird}, 
and which physical influence they may be expected to have there. 

Before proceeding to the scalar field case, let us note that
the case of ``normal'' hydrodynamic fluctuations is for completeness
briefly reviewed in appendix~A.

%
\section{Hydrodynamics coupled to a scalar field}

%
\subsection{Original setup}

We start by recapitulating the basic equations 
{\em without fluctuations}. 
For generality the system is put in a curved background 
with a metric~$g^{\mu\nu}_{ }$, even if for some 
considerations it is sufficient to subsequently restrict
to the Minkowski metric  
or to linear perturbations around it. 

As the basic variables, we adopt a real scalar field $\phi$
and a plasma parametrized by a local temperature $T$ and a flow 
velocity $u^{\mu}_{ }$. The energy-momentum tensor is postulated to 
have the form 
\ba
 T^{\mu\nu}_{  }
   & \equiv & 
 \phi^{,\mu}_{ }\phi^{,\nu}_{ }
 - \frac{ g^{\mu\nu}_{ }  \phi^{ }_{,\alpha} \phi^{,\alpha}_{ } }{2} 
 + w \, u^{\mu}_{ }u^{\nu}_{ }+ p \, g^{\mu\nu}_{ }
 \;, 
 \\
  p & \equiv & p^{ }_0(T) - V(\phi,T)
 \;, 
 \quad
 w \; \equiv \; T \partial^{ }_T p
 \;, \la{Tmunu}
\ea
where
$
 p 
$
is the pressure,  
$
 w 
$
is the enthalpy density, and 
$u^{\mu}_{ }$  is the flow velocity. 
By $()^{ }_{,\mu}$ we denote a partial derivative 
in the $x^\mu_{ }$-direction, whereas $()^{ }_{;\mu}$ is 
a covariant derivative.  
For $g^{\mu\nu}_{ }$
we employ the ``mostly plus'' metric convention, 
so that $u^{ }_\mu u^{\mu}_{ } = - 1$.  
Within a derivative expansion (slow variations) 
the basic equations are~\cite{ikkl}
\ba
 T^{\mu\nu}_{;\mu}
 &  = & 0 \;,
 \la{dTmunu} \\ 
 \phi^{;\mu}_{;\mu} - \gamma \, u^{\mu}_{ } \phi^{ }_{,\mu}
 - \partial^{ }_{\phi} V 
 & = & 0
 \;. \la{dphi}
\ea
The coefficient $\gamma$ parametrizes entropy production in regions
where the scalar field varies (i.e.\ particularly around bubble walls):  
$
 T(s u^{\mu}_{ })^{ }_{;\mu} = \gamma (u^{\mu}_{ }\phi^{ }_{,\mu})^2
$, 
where $s \equiv \partial^{ }_T p$.

Without the scalar field contribution, the energy-momentum tensor would
be that of an ideal fluid. In that system phase transition fronts and shocks 
appear as discontinuities~\cite{landau6}. Originally, the introduction
of $\phi$ was motivated by having a microscopically 
adjustable parametrization of the entropy production that 
takes place at these discontinuities~\cite{ikkl}. 
However, lately the same model is also used for studying the 
subsequent stages with more complicated dynamics~\cite{weird}. 
With many overlapping sound waves, the system starts 
to resemble a thermal plasma with various
random motions taking place simultaneously. 

We note in passing that we do 
not consider here the microscopic origin of the coefficient~$\gamma$.
In general it is a function of $\phi$, though 
it is expected to have a non-zero value even as $\phi\to 0$~\cite{dietrich}.
In the following we are concerned with the ``final state'' of the system, 
which in the context of the electroweak phase transition means the 
low-temperature Higgs phase, $\phi \approx \phi^{ }_0(T)$. 
We 
shift $\phi$ by $ \phi^{ }_0(T) $ so that after the shift
$V(\phi,T) \equiv \fr12 m^2(T) \phi^2$,
and  
ignore scalar field self-interactions. Terms originating
from the shift by $ \phi^{ }_0(T) $ 
have been included in $p^{ }_0(T)$. 

The equations presented above 
should apply in the so-called hydrodynamic
regime~\cite{landau9}, i.e.\ at time and length scales
$\gsim 1/ (\alpha^2 T)$, where $\alpha$ is a coupling
characterizing the interactions within the plasma. 
At the electroweak epoch $T\sim 100$~GeV, and the Hubble radius 
is $H^{-1} \sim 10^{15} T^{-1}$. The bubble distance
scale is a macroscopic fraction of the latter, say 
$10^{-2} H^{-1}$~\cite{weird}, and thus indeed huge
compared with $1/(\alpha^2 T)$, 
even if $\alpha \sim 10^{-2}$. 

%
\subsection{Thermal fluctuations}
\la{ss:fluct}

Whenever dissipation is present, 
the fluctuation-dissipation theorem needs to 
be respected~\cite{landau9}. This implies that the scalar field  
equation in \eq\nr{dphi} should actually be corrected into 
\be
 \phi^{;\mu}_{;\mu} - \gamma \, u^{\mu}_{ } \phi^{ }_{,\mu}
 - m^2 \phi 
 + \xi 
 \; = \; 0
 \;, \la{phi_eq}
\ee
where $\xi$ is a stochastic noise term. 
The autocorrelator of the noise is assumed to take the form
\be
 \bigl\langle \xi(\mathcal{X}) \xi(\mathcal{Y}) \bigr\rangle \; = \; 
 \frac{\Omega\, \delta(\mathcal{X-Y})}{\sqrt{-\det g^{ }_{\mu\nu}}} 
 \;, \quad 
 \mathcal{X} \; = \; (t,\vec{x})
 \;, \la{noise}
\ee
where $\Omega$ is a coefficient whose value is determined presently
(cf.\ \eq\nr{Omega}). 

Let us solve \eq\nr{phi_eq} in local Minkowskian coordinates 
($g^{ }_{\mu\nu} \to \eta^{ }_{\mu\nu} = (\mbox{$-$+++})$) in 
a domain of a constant 4-velocity $u^{\mu}_{ }$. Considering 
times $ \gg \gamma^{-1}_{ }$ so that 
initial transients have died out, the solution can be written as
\ba
 \phi(\mathcal{X}) & = & \int_{\mathcal{Y}} 
 G^{ }_\rmii{R}(\mathcal{X-Y}) \, \xi(\mathcal{Y}) \;, \\
 G^{ }_\rmii{R}(\mathcal{X})  & = &
 \int_{\mathcal{P}} \frac{e^{i \mathcal{P}\cdot\mathcal{X}}}
 {\mathcal{P}^2 + i \gamma \mathcal{P}\cdot u + m^2}  
 \;, \la{G_R}
\ea
where
$
 \mathcal{P} = (\omega,\vec{p})
$
and
$
 \int_{\mathcal{P}} \equiv \int_{-\infty}^{\infty} \! 
 \frac{{\rm d}\omega}{2\pi} \int\! \frac{{\rm d}^d\vec{p}}{(2\pi)^d} 
$. 
For $\gamma > 0$ the poles in \eq\nr{G_R} are in the 
lower half-plane, and therefore $G^{ }_\rmii{R}(\mathcal{X})$ 
is a retarded Green's function.
Making use of the autocorrelator
in \eq\nr{noise} the 2-point function becomes
\ba
 \bigl\langle \phi(\mathcal{X}) \phi(\mathcal{Y})\bigr\rangle
 & = &  
 \Omega \int_{\mathcal{Z}} G^{ }_\rmii{R}(\mathcal{X-Z}) 
 G^{ }_\rmii{R}(\mathcal{Y-Z})
 \nn 
 & = & 
 \int_{\mathcal{P}} \frac{\Omega\, 
 e^{i \mathcal{P}\cdot(\mathcal{X-Y})}}
 {(\mathcal{P}^2 + m^2)^2 + 
  \gamma^2 (\mathcal{P}\cdot u)^2}
 \;. \la{phi_phi_1} 
\ea
We thus see that, 
in analogy with the real-time formalism of thermal field theory~\cite{sch},
Feynman rules for this system 
contain two types of propagators, 
the retarded propagator $G^{ }_\rmii{R}$ 
and a statistical propagator $\langle \phi \phi \rangle$ originating from 
the average $G^{ }_\rmii{R} \langle \xi \xi \rangle G^{ }_\rmii{R}$.

The integral over $\omega$ can be carried out in \eq\nr{phi_phi_1}. 
In particular, setting the time arguments equal and 
denoting $\epsilon_p^2 \equiv p^2 + m^2$ 
with $p \equiv |\vec{p}|$, we get
\ba
 && \hspace*{-1.0cm}
 \bigl\langle
  \phi(t,\vec{x})
  \phi(t,\vec{y})
 \bigr\rangle
 \; = \; 
 \frac{\Omega}{2\gamma}
 \int_{\vec{p}}
 \frac{e^{i \vec{p}\cdot(\vec{x-y})}}{2\epsilon^{ }_p}
 \nn 
 & \times & 
 \biggl\{
   \frac{1}{\epsilon^{ }_p u^{0}_{ } - \vec{p}\cdot\vec{u}} 
 + 
   \frac{1}{\epsilon^{ }_p u^{0}_{ } + \vec{p}\cdot\vec{u}} 
 \biggr\}
 \;. \la{phi_phi_2}
\ea

In order to fix the value of $\Omega$, 
let us compare \eq\nr{phi_phi_2} with the direct computation
of the 2-point correlator in an ensemble 
defined by the density matrix
$
 \hat{\rho} \equiv {Z}^{-1}_{ } e^{-(\hat{H} u^0 - \hat{K}^i_{ } u^i_{ })/T}
$, 
where $Z$ is the partition function, 
$\hat{H}$ is the Hamiltonian, and $\hat{K}^i_{ }$ is the momentum
operator. We obtain
\ba
 & & \hspace*{-1cm}
 \tr \bigl\{\hat{\rho}\, \hat{\phi}(t,\vec{x}) \hat{\phi}(t,\vec{y}) \bigr\}
 \; = \; 
 \int_{\vec{p}}
 \frac{e^{i \vec{p}\cdot(\vec{x-y})}}{2\epsilon^{ }_p}
 \nn  & \times & 
 \bigl\{
 1 + \nB^{ }(\epsilon^{ }_p u^{0}_{ } - \vec{p}\cdot\vec{u}) 
   + \nB^{ }(\epsilon^{ }_p u^{0}_{ } + \vec{p}\cdot\vec{u}) 
 \bigr\}
 \;, \hspace*{4mm} \la{phi_phi_3} 
\ea 
where $\nB^{ }(x) \equiv 1 / [\exp(x/T) - 1]$ 
is the Bose distribution. The hydrodynamic description of 
\eq\nr{phi_phi_2} is supposed 
to apply for $\epsilon^{ }_p,p \ll T$. Expanding 
$\nB^{ }(x) \approx T/x$ and comparing
\eqs\nr{phi_phi_2} and \nr{phi_phi_3}, we uniquely identify 
the noise autocorrelator $\Omega$ as 
\be
 \Omega \; = \; 2 \gamma T 
 \;. \la{Omega}
\ee
More generally, if $\phi$ is not in equilibrium with the medium, 
the $T$ in \eq\nr{Omega} could differ from that in \eq\nr{Tmunu}. 

%
\subsection{Energy-momentum correlator}

Let us now define a ``transverse-traceless'' (TT)
correlator of the energy-momentum tensor $T^{\mu\nu}_{ }$, 
after choosing the spatial momentum to point in the $z$-direction: 
\be
 C^\rmii{TT}_{\Delta}(k_{ }^0,k) \; \equiv \; 
 \int_{\mathcal{X}} e^{i k_{ }^0 t- i k z}
 \, \Bigl\langle
 \fr12
 \bigl\{ 
   T^{xy}_{ }(t,\vec{x}) \,,\,
   T^{xy}_{ }(0)
 \bigr\}
 \Bigr\rangle
 \;. \la{TT}
\ee
With the classical fields that appear in hydrodynamics, 
operator ordering plays actually no role.  
For convenience we denote the infrared limit of this correlator by 
\be
 \lim_{k_{ }^0,k\to 0} C^\rmii{TT}_{\Delta}(k_{ }^0,k)
 \; \equiv \; 2 \eta T
 \;. \la{eta}
\ee
Through a standard Kubo relation, $\eta$ can be interpreted 
as an effective overall ``shear viscosity'' of the coupled system
(fluid + $\phi$), but for the purposes of the present paper
\eq\nr{eta} can equally well be taken as a definition of $\eta$. 
In any case, the infrared contribution to 
the differential production rate of the energy density carried by 
gravitational waves reads~\cite{jacopo}
\be
 \lim_{k\to 0}
 \frac{{\rm d}e^{ }_\rmii{GW}}{{\rm d}t\, {\rm d}^3\vec{k}}
 \; = \;
 \frac{4\eta T}{\pi^2 m_\rmi{Pl}^2}
 \;,  \la{eGW}
\ee
where $m^{ }_\rmi{Pl} = 1.22\times 10^{19}$~GeV is the Planck mass. 

Rather than directly computing the correlator in \eq\nr{TT}, which could
be achieved through the use of \eq\nr{phi_phi_1}, it is 
illuminating to couple the system to a metric perturbation 
$
 h^{ }_{xy}(t,z) \; \equiv \; h^{ }_{xy} \, e^{-i k_{ }^0 t + i k z}
$. 
The response of the 
expectation value of $T^{xy}_{ }$ to this background yields 
the retarded correlator~\cite{guy}, from which the time ordering 
in \eq\nr{TT} can be readily extracted 
(assuming that the scalar field is in thermal equilibrium 
{\em \`a la} \eq\nr{Omega}): 
\be
 C^\rmii{TT}_{\Delta}(k_{ }^0,k) \; = \; 
 \bigl[ 1 + 2 \nB^{ }(k_{ }^0) \bigr]
 \, \lim_{h^{ }_{xy} \to 0}
 \im \bigl[ 
 \delta
 \bigl\langle 
   T^{xy}_{ }
 \bigr\rangle / \delta h^{ }_{xy}
 \bigr]
 \;. \la{kms}
\ee

Working to leading order in small perturbations ($\phi, u^i_{ }$)
and to linear order in $h^{ }_{xy}$, \eq\nr{phi_eq} takes the form 
\be
 \bigl( \partial_t^2 + \gamma \partial^{ }_t - \nabla^2 + m^2 \bigr) \phi 
 \; = \; 
 \xi - 2 h^{ }_{xy} \phi^{ }_{,x,y} 
 \; + \; 
 \rmO(h_{xy}^2,\phi\, u^i_{ })
 \;. 
\ee
This can be solved as 
\ba
 \phi(\mathcal{X})
 & \approx &
 \int_{\mathcal{Y}} G^{ }_\rmii{R}(\mathcal{X-Y})\, \xi(\mathcal{Y})
 \\ 
 & - &  
 2 \int_{\mathcal{Y,Z}} 
 G^{ }_\rmii{R}(\mathcal{X-Y}) \,
 h^{ }_{xy}(\mathcal{Y}) \,
 G^{ }_{\rmii{R},x,y}(\mathcal{Y-Z}) \,
 \xi(\mathcal{Z})
 \;. \nonumber 
\ea
Inserting into 
$
 \bigl\langle T^{xy}_{\phi}
 \bigr\rangle^{ }_{ }
 \; \equiv \; 
 \bigl\langle
  \phi^{ }_{,x}
  \phi^{ }_{,y}
 \bigr\rangle^{ }_{ }
$, 
averaging over fluctuations, integrating over energy,
and omitting terms suppressed by $k^2/\gamma^2$, 
we obtain
\ba
 && \hspace*{-8mm}
 \bigl\langle T^{xy}_{\phi}\bigr\rangle^ { }_{ }
 \; \approx \; 
 T h^{ }_{xy}(t,z) 
 \int_{\vec{p}} \frac{p_x^2 p_y^2}{\epsilon_p^4} 
 \frac
 {\textstyle 1 + \frac{i \gamma}{k_{ }^0 + i \gamma} } 
 {\textstyle 1 - \frac{k^0_{ }(k^0_{ }+ 2 i \gamma)}{4 \epsilon_p^2} }
 \;. \hspace*{3mm} \la{result} 
\ea
For fixed $k^0_{ }/\gamma$ and to leading order in 
$\gamma^2 / \epsilon^{2}_p$, 
we thus find a Lorentzian shape
$
 1 + 
 \frac{i \gamma}{k_{ }^0 + i \gamma}
$
for the retarded correlator.

%
\subsection{Ultraviolet problem and finite part}

Inserting \eq\nr{result} into \eqs\nr{eta} and \nr{kms} yields 
a scalar contribution to the effective shear viscosity, 
\be
 \delta \eta = 
 \frac{T}{\gamma} \int_{\vec{p}} \frac{p_x^2 p_y^2}{(p^2 + m^2)^2}
 \biggl( 
   1 + \frac{\gamma^2}{p^2 + m^2}
 \biggr)
 \;. \la{eta_1}
\ee
The same result can be obtained from 
a direct computation of the correlator in \eq\nr{TT}, along 
lines illustrated for normal hydrodynamic fluctuations in 
appendix~A. 

As is familiar from classical field theory~\cite{mclerran,arnold}, 
the result is power-divergent at large momenta. Cutting off
large momenta so that $p \le \Lambda$, the divergent part reads
\be
 \delta \eta |^{ }_\rmi{div} =  \frac{ T }{90 \pi^2 \gamma}
 \bigl[ \Lambda^3 + 3 (\gamma^2 - 2 m^2) \Lambda\bigr]
 \;. \la{eta_2}
\ee
If the theory is rather regularized on a (comoving) lattice, 
so that the autocorrelator in \eq\nr{noise} becomes 
$
 \Omega\, \delta(\mathcal{X-Y})/\sqrt{-\det g^{ }_{\mu\nu}} \to 
 \Omega\, \delta^{ }_{x^0_{ },y^0_{ }}\delta^{ }_{\vec{x},\vec{y}} / 
 (a^{ }_t a_s^3)
$, 
where $a^{ }_t$ and $a^{ }_s$ are the temporal and spatial lattice
spacings, respectively, 
partial integration and rotational invariance
permit to reduce the cubic divergence to a known tadpole~\cite{lat1,lat2}, 
\ba
 && 
 \delta \eta |^{ }_\rmi{lat} \; 
    \stackrel{\frac{1}{a^{ }_{\! s}} \gg\, m,\, \gamma }{\approx} 
 \;  \frac{T}{\gamma}
 \int_{\vec{p}} \frac{\tilde{p}^2_x \tilde{p}^2_y }{\tilde{p}^4}
 \; = \; \frac{T \beta }{3 \gamma a_s^3}
 \;, \la{eta_lat} \\
 && 
 \beta \; \equiv \; 
 \int_{\vec{p}} \frac{a_s^3}{4 \sum_i \sin^2(\frac{a^{ }_s p^{ }_i}{2})}
 \; = \; 
 \Gamma^2\Bigl(\fr1{24}\Bigr)
 \Gamma^2\Bigl(\fr{11}{24}\Bigr)
 \frac{\sqrt{3}-1}{ 192 \pi^3}
 \;, \nn \la{eta_lat2}
\ea
where $\tilde{p}^{ }_i \equiv 
\frac{2}{a^{ }_s}\sin(\frac{a^{ }_s p^{ }_i}{2})$ 
are lattice momenta and the integration
is carried out over the first Brillouin zone. 

The integral in \eq\nr{eta_1} also has a finite part, which can be determined 
with dimensional regularization in $d$ spatial dimensions: 
\be
 \delta \eta |^{ }_\rmi{fin} = 
 \frac{ T m^d \Gamma(2-{d}/{2}) }
 { \gamma (4\pi)^{{d}/{2}} d(d-2)  }
 \biggl( 1 - \frac{\gamma^2 d}{4 m^2} \biggr)
 \;. \la{eta_3}
\ee
Setting $d=3$ and $\gamma \sim m \sim \alpha^2 T$, so that we are safely 
in the hydrodynamic regime, this is parametrically 
a very subleading contribution, 
$\delta \eta |^{ }_\rmii{fin}  \ll T^3/\alpha^2$. 
(We note that \eq\nr{eta_3} becomes negative in the overdamped
regime $\gamma \gsim m$, but this is of no concern,
given that the full result from \eq\nr{eta_1} remains positive.) 

That \eqs\nr{eta_1}--\nr{eta_3} diverge as $\gamma \ll m$, 
is familiar from other weakly coupled systems~\cite{jeon} and from 
the contribution of hydrodynamic fluctuations~\cite{kovtun}. 


%
\section{Conclusions} 

It seems conceptually attractive to incorporate 
scalar fluctuations into hydrodynamic simulations of 
cosmological phase transition dynamics. 
To begin with, this is theoretically necessary for
respecting the fluctuation-dissipation theorem at a late
time when the order parameter is approaching thermal equilibrium. 
In addition, thermal fluctuations would {\em in principle} lead to 
automatic bubble nucleations, even if in practice 
multicanonical simulations are needed for studying these rare
events with their proper weights~\cite{mr1}. 
Fluctuations may also induce a first order phase transition~\cite{classic},
even if this would not happen with a scalar field alone.
Finally, fluctuations would help in probing the 
instability of growing bubbles~\cite{Huet}. 

The practical inclusion of hydrodynamic fluctuations leads to 
powerlike ultraviolet divergences. In the regime of linear
perturbations, the contribution from scalar
fluctuations is cubically divergent in the formal
continuum limit (cf.\ \eq\nr{eta_2}), whereas
that from normal hydrodynamic fluctuations is linearly divergent
(cf.\ \eq\nr{eta_hydro}).
There is perturbative evidence that 
a cutoff-independent framework may be obtained by  
treating shear and bulk viscosities as 
``bare'' parameters,  
and introducing counterterms for all possible 
thermodynamic functions, even if this 
leads to a rather complicated framework
(cf.\ ref.~\cite{akamatsu} for recent work and references). 
On a lattice, the loss of rotational symmetry 
may also become a concern~\cite{mclerran,arnold}.  
However, turning the tables, a ``bare simulation'' would
yield a well-predicted shear viscosity, cf.\ \eq\nr{eta_lat}, 
and a corresponding contribution to the differential gravitational wave
production rate at late times, cf.\ \eq\nr{eGW}. 
The amplitude of this component can be tuned at
will by changing the lattice spacing $a^{ }_s$ or  
the amplitude $\Omega$ of the noise auto-correlator. 
If the value of $\Omega$ deviates from that in \eq\nr{Omega}, 
the resulting $\eta$ scales as $\Omega^2/(2\gamma T)^2$
relative to \eq\nr{eta_lat}. This behaviour of the overall magnitude, 
together with a corresponding spectral shape, are worth testing 
as a clean calibration of the measurement algorithm.

%
\section*{Acknowledgements}

This work was supported by the Swiss National Science Foundation
(SNF) under grant 200020-168988. 

%
\appendix
\renewcommand{\thesection}{Appendix~\Alph{section}:}
\renewcommand{\thesubsection}{\Alph{section}.\arabic{subsection}}
\renewcommand{\theequation}{\Alph{section}.\arabic{equation}}

%
\section{Contribution from normal hydrodynamic fluctuations}

For completeness and comparison with the scalar field case, 
we review here the contribution of normal hydrodynamic
fluctuations to shear viscosity~\cite{kovtun,akamatsu}.

In terms of the fundamental theory, we are considering 
a density matrix parametrized by a flow velocity $u^\mu_{ }$
and a temperature $T$: 
\be
 \hat\rho \; \equiv \; 
 \frac{1}{Z}
 \exp
 \biggl( 
  - \frac{\hat{H} u^{0}_{ }- \hat{K}^i_{ } u^i_{ }}{T}
 \biggr) 
 \;, 
 \la{rho}
\ee
where
$
 \hat{H} 
$
is the Hamiltonian,  
$
 \hat{K}^i_{ }
$
is the momentum operator, 
$u^{ }_\mu u^{\mu}_{ } = - 1$, 
and $Z$ is chosen so that 
$
 \tr\hat{\rho} = 1 
$. 
Factoring out $u^0_{ }/T$, 
the thermodynamic pressure is defined as
\be
 p\biggl( \frac{T}{u^0_{ }},\frac{\vec{u}}{u^0_{ }} \biggr)
 \; \equiv \; 
 \lim_{V\to\infty} \frac{T \ln Z }{u^0_{ }V} 
 \;, 
\ee
where $V$ is the volume. 
Poincar\'e invariance implies that
(cf.\ ref.~\cite{meyer} and references therein)
\be
 p\biggl( \frac{T}{u^0_{ }},\frac{\vec{u}}{u^0_{ }} \biggr)
 \; = \; p(T,\vec{0}) \; \equiv \; p(T)  
 \;. \la{poincare}
\ee
{}From this relation it can be shown that 
\ba
 \lim_{V\to\infty} \frac{\langle\hat{H}\rangle}{V}
 & = & T\partial^{ }_T p  - p \; \equiv \; e
 \;, \\
 \lim_{V\to\infty} \frac{\langle \hat{K}^{i}_{ }\rangle }{V} 
 & = &
 u^0_{ }u^{i}_{ } T \partial^{ }_T p \; \equiv \; u^0_{ }u^{i}_{ } w 
 \;, 
\ea
where $e$ is the energy and $w$ the enthalpy density. 
These expectation values appear as parts of 
$
 T^{\mu\nu}_\rmi{ideal} = p g^{\mu\nu}_{ } + w u^{\mu}_{ }u^{\nu}_{ }
$.
The ``susceptibility'' related to $\hat{K}^{i}_{ }$ 
becomes
\be
 \left. \lim_{V\to\infty} \frac{\langle \hat{K}^{i}_{ } \hat{K}^{j}_{ }
 \rangle}{V} \right|^{ }_{u^i = 0}
 \; = \; T w\, \delta^{ij}_{ }
 \;. \la{susc_hydro}
\ee

Let us now assume that $T$ and $u^\mu_{ }$ 
are not constant but vary slowly, and 
expand the expectation value of the energy-momentum tensor 
to first order in gradients. Following ref.~\cite{landau9}, 
small variations cannot be distinguished from 
occasional long-wavelength thermal fluctuations ($S^{\mu\nu}_{ }$), 
which must therefore be added as ingredients: 
\ba
 T^{\mu\nu}_{ } & = & 
 T^{\mu\nu}_\rmi{ideal}
 - \eta \Delta^{\mu\rho}_{ }\Delta^{\nu\sigma}_{ }
 \Bigl(
  u^{ }_{\rho;\sigma} + u^{ }_{\sigma;\rho}
 - \frac{2 g^{ }_{\rho\sigma}}{d}  u^{\gamma}_{;\gamma}  
 \Bigr) 
 \nn
 & - & \zeta \Delta^{\mu\nu}_{ } u^{\gamma}_{;\gamma}
 + S^{\mu\nu}_{ }
 \;, \la{Tmunu_hydro} 
\ea
where 
$
 \Delta^{\mu\nu}_{ } \;\equiv\;
 g^{\mu\nu}_{ }+ u^{\mu}_{ }u^{\nu}_{ } 
$
is a projector onto directions orthogonal to $u^{\mu}_{ }$, 
and $\eta,\zeta$ are the shear and bulk viscosities.
The noise correlator takes the form~\cite{landau9,kapusta}
\ba
 && \hspace*{-1.5cm}
 \bigl\langle 
  S^{\mu\nu}_{ }(\mathcal{X})
  S^{\rho\sigma}_{ }(\mathcal{Y})
 \bigr\rangle
 \; = \; 2 T 
 \Bigl[
  \eta \, \bigl( \Delta^{\mu\rho}_{ } \Delta^{\nu\sigma}_{ }
  + \Delta^{\mu\sigma}_{ } \Delta^{\nu\rho}_{ } \bigr)
 \nn 
 & + &
    \Bigl( \zeta - \frac{2\eta}{d} \Bigr)
    \Delta^{\mu\nu}_{ } \Delta^{\rho\sigma}_{ }
 \Bigr] \frac{\delta_{ }(\mathcal{X-Y})}
 {\sqrt{- \det g^{ }_{\mu\nu}}}
 \;. \hspace*{5mm} \la{noise_hydro}
\ea 

We now restrict ourselves to flat spacetime and 
consider small perturbations of $T$ und $u^i_{ }$
around the equilibrium values $T^{ }_0$ and $0$, respectively. 
Small (non-relativistic) velocity fluctuations are denoted by $v^i_{ }$, 
and equilibrium values by 
$p^{ }_0, e^{ }_0, w^{ }_0, \eta^{ }_0, \zeta^{ }_0$, etc. 
It is helpful to go over into Fourier space, 
$
 f(\mathcal{X}) = \int_{\mathcal{P}} 
 e^{i\mathcal{P}\cdot\mathcal{X}} 
 f(\mathcal{P})
$, 
where 
$\mathcal{P} \equiv (\omega,\vec{p})$, 
and we also define $p \equiv |\vec{p}|$.
Putting the terms originating from $S^{\mu\nu}_{ }$ on the right-hand side of 
the equation, defining
\be
 \xi^\mu_{ }(\mathcal{P})
 \;\equiv \; \frac{\mathcal{P}^{ }_{\nu} S^{\mu\nu}_{ }(\mathcal{P})}
 {w^{ }_0}
 \;, \la{xi_def}
\ee
and denoting 
\ba
 \theta & \equiv & \ln \biggl( \frac{T}{T^{ }_0} \biggr)
 \;, \quad
 c_s^2 \; \equiv \; 
 \frac{\partial p^{ }_0}{\partial e^{ }_0}
 \;, \\ 
 \bar{\eta}^{ }_1 & \equiv & \frac{\zeta^{ }_0 +
 \frac{2(d-1)}{d} \eta^{ }_0}{w^{ }_0}
 \;, \quad 
 \bar{\eta}^{ }_2 \; \equiv \; \frac{\eta^{ }_0}{w^{ }_0}
 \;, 
\ea
energy-momentum conservation
$\partial^{ }_{\mu}T^{\mu\nu}_{ }= 0$
implies that 
the temperature and velocity fluctuations are given by 
\ba 
 \theta(\mathcal{P}) & = & 
 \frac{c_s^2
 \bigl[ 
   (\omega+ i \bar{\eta}^{ }_1 p^2) \xi^0_{ }(\mathcal{P}) 
   + p^i_{ } \xi^i_{ }(\mathcal{P})
 \bigr]}
 {\omega^2 + i \bar{\eta}^{ }_1 \omega p^2 - c_s^2 p^2}
 \;, \\ 
 v^i_{ }(\mathcal{P}) & = & 
 \frac{p^i_{ }
 \bigl[ 
   c_s^2 \xi^0_{ }(\mathcal{P}) 
   + \frac{\omega p^j_{ }}{p^2}\, \xi^j_{ }(\mathcal{P})
 \bigr]}
 {\omega^2 + i \bar{\eta}^{ }_1 \omega p^2 - c_s^2 p^2}
 \; + \; 
 \frac{
 \bigl( \delta^{ij}_{ } - \frac{p^i_{ }p^j_{ }}{p^2} \bigr)
 \,\xi^j_{ }(\mathcal{P})
 }{\omega + i \bar{\eta}^{ }_2 p^2}
 \;. \nn
\ea
These are the analogues of \eq\nr{G_R}. 
Inserting \eq\nr{xi_def}, evaluating the thermal average
according to \eq\nr{noise_hydro}, and noting that to leading order in 
$v^i_{ }$ only $S^{ij}_{ }$ contributes, the 
velocity correlator becomes (cf.\ e.g.\ ref.~\cite{yaffe})
\ba
 \bigl\langle v^i_{ }(\mathcal{P}) v^{j}_{ }(\mathcal{Q}) \bigr\rangle
 & = & 
 \delta(\mathcal{P+Q}) \, G^{ij}_{ }(\mathcal{P})
 \;, \la{Gij} \\ 
 G^{ij}_{ }(\mathcal{P})
 & = & 
 \frac{2T^{ }_0}{w^{ }_0}
 \biggl[ 
 \frac{\bar{\eta}^{ }_1 \omega^2 p^i_{ } p^j_{ }}
 {(\omega^2 - c_s^2 p^2)^2 + \bar{\eta}_1^2 \omega^2 p^4 }
 \nn 
 & + & 
 \frac{
 \bar{\eta}^{ }_2 
 \bigl( p^2 \delta^{ij}_{ } - p^i_{ }p^j_{ } \bigr)
 }{\omega^2 + \bar{\eta}_2^2 p^4}
 \biggr]
 \;. \la{v_v}
\ea
Here $\delta(\mathcal{P+Q}) \equiv (2\pi)^D_{ }\delta^{(D)}_{ }(\mathcal{P+Q})$
and $D \equiv d+1$. These are the analogues of \eq\nr{phi_phi_1}.

It is useful to crosscheck that 
\eq\nr{v_v} reproduces 
the susceptibility from \eq\nr{susc_hydro}. To leading order 
in velocities, the components $S^{0i}_{ }$ have a vanishing 
correlator. Therefore, from \eq\nr{Tmunu_hydro},  
$
 T^{0i}_{ } \simeq w^{ }_0 v^i_{ }
$,
and 
$
 \bigl\langle T^{0i}_{ }(\mathcal{P}) T^{0j}_{ }(\mathcal{Q}) \bigr\rangle 
$
is directly proportional to \eq\nr{v_v}. The equal-time correlator
relevant for \eq\nr{susc_hydro}
can be obtained by integrating \eq\nr{v_v}
over the frequency, $\int \frac{{\rm d} \omega}{2\pi}$.
Thereby we reproduce \eq\nr{susc_hydro} in $d$ spatial dimensions.

We now move on to the correlator in \eq\nr{TT}. From 
\eq\nr{Tmunu_hydro},  
\be
 T^{xy}_{ } \approx w^{ }_0 v^x_{ }v^y_{ } -
 \eta^{ }_0
 \bigl(
  \partial^x_{ }v^y_{ } + \partial^y_{ }v^x_{ }
 \bigr)
 + S^{xy}_{ }
 \;. 
\ee
According to \eq\nr{noise_hydro}, the noise part gives 
$
 \bigl\langle
    S^{xy}_{ }(\mathcal{K}) S^{xy}_{ }(\mathcal{Q})
 \bigr\rangle
 = 2 \eta^{ }_0 T^{ }_0 \delta(\mathcal{K+Q})
$, 
as expected from \eq\nr{eta}.
In momentum space,  
the middle term vanishes for 
$\mathcal{K}=(k_{ }^0,k\,\vec{e}^{ }_z)$ 
as is relevant for \eq\nr{TT}.
Following ref.~\cite{kovtun},  
we consider the 1-loop contribution 
from $w^{ }_0 v^x_{ }v^y_{ }$:
\ba
 && \hspace*{-0.5cm} 2 T^{ }_0\, \delta \eta^{ }_0 
 \\ 
 & \equiv & 
 \frac{w_0^2}{\delta(\mathcal{K+Q})}
 \int_{\mathcal{P,R}}
 \bigl\langle
  v^x_{ }(\mathcal{P}) v^y_{ }(\mathcal{K-P}) 
  v^x_{ }(\mathcal{R}) v^y_{ }(\mathcal{Q-R}) 
 \bigr\rangle 
 \nn 
 & = & 
 w_0^2 \, \int_{\mathcal{P}}
 \Bigl[
   G^{xx}_{ }(\mathcal{P}) G^{yy}_{ }(\mathcal{K-P})
 + \, G^{xy}_{ }(\mathcal{P}) G^{yx}_{ }(\mathcal{K-P})
 \Bigr] 
 \;, \nonumber
\ea
where we inserted \eq\nr{Gij}. Substituting 
\eq\nr{v_v}, integrating over $\omega$, and setting 
$\mathcal{K}\to 0$ as is sufficient according to \eq\nr{eta},  
we obtain
\ba
 \lim_{\mathcal{K}\to 0}\delta \eta^{ }_0 & = &  
 2 T^{ }_0 \int_{\vec{p}}
 \biggl\{ 
   \frac{p_x^2 p_y^2}{2 \bar{\eta}^{ }_1 p^6}
  + \frac{p^4 - 2 p^2 p_x^2 + 2 p_x^2 p_y^2}{4\bar{\eta}^{ }_2 p^6}
 \nn & & \hspace*{1cm}
  + \, \frac{\bar{\eta}^{ }_2 (p^2 p_x^2 - 2 p_x^2p_y^2)}
    {\bar{\eta}^{ }_2(\bar{\eta}^{ }_1+\bar{\eta}^{ }_2)p^6 + c_s^2 p^4}
 \biggr\}
 \nn 
 & = & 
 \frac{T^{ }_0}{d+2} 
 \int_{\vec{p}}
 \biggl\{
   \frac{1}{ \bar{\eta}^{ }_1 p^2 d} 
 + \frac{d^2 - 2}{2 \bar{\eta}^{ }_2 p^2 d}
 \nn & & \hspace*{1cm}
 + \, \frac{2 \bar{\eta}^{ }_2}
{ \bar{\eta}^{ }_2(\bar{\eta}^{ }_1+\bar{\eta}^{ }_2)p^2 + c_s^2 } 
 \biggr\}
 \;, \la{eta_hydro}
\ea
where we made use of rotational symmetry to write
$
 \int_{\vec{p}} p^{ }_i p^{ }_j p^{ }_k p^{ }_l \phi(p^2)
 = 
 \frac{\delta^{ }_{ij}\delta^{ }_{kl} + \delta^{ }_{ik}\delta^{ }_{jl}
 + \delta^{ }_{il}\delta^{ }_{jk}}{d(d+2)} 
 \int_{\vec{p}} p^4 \phi(p^2)
$
and 
$
 \int_{\vec{p}} p^{ }_i p^{ }_j \phi(p^2) 
 = 
 \frac{\delta^{ }_{ij}}{d}
 \int_{\vec{p}} p^2 \phi(p^2)
$. 
Upon setting $d=3$,  
$\int_{\vec{p}} \frac{1}{p^2} = \frac{\Lambda}{2\pi^2}$,
and omitting the last term, this agrees with 
refs.~\cite{kovtun,akamatsu}.
The last term was omitted because at small $p$ it is suppressed
by $\sim\bar{\eta}_i^2 p^2 / c_s^2$ compared with the other terms. 
In the continuum limit of a hydrodynamic simulation 
it should, however, be included as
$ 2 / [(\bar{\eta}^{ }_1 + \bar{\eta}^{ }_2) p^2]
$.  

\small{
%

}


\begin{thebibliography}{99}

\bibitem{lisa_transition}
  C.~Caprini {\it et al.},
  {\it Science with the space-based interferometer eLISA. 
  II: Gravitational waves from cosmological phase transitions,}
  JCAP {04} (2016) 001
  [1512.06239].

\bibitem{weird}
  M.~Hindmarsh, S.J.~Huber, K.~Rummukainen and D.J.~Weir,
  {\it Shape of the acoustic gravitational wave power spectrum from
  a first order phase transition,}
  Phys.\ Rev.\ D {96} (2017) 103520
  [1704.05871].

\bibitem{jacopo}
  J.~Ghiglieri and M.~Laine,
  {\it Gravitational wave background from Standard Model physics:
  Qualitative features,}
  JCAP {07} (2015) 022
  [1504.02569].

\bibitem{db_gdm}
  D.~B\"odeker and G.D.~Moore,
  {\it Electroweak Bubble Wall Speed Limit,}
  JCAP {05} (2017) 025
  [1703.08215].

\bibitem{ikkl}
  J.~Ignatius, K.~Kajantie, H.~Kurki-Suonio and M.~Laine,
  {\it The growth of bubbles in cosmological phase transitions,}
  Phys.\ Rev.\ D {49} (1994) 3854
  [astro-ph/9309059].

\bibitem{landau6}
  L.D.~Landau and E.M.~Lifshitz,
  {\it Fluid Mechanics}
  (Butterworth-Heinemann, Oxford, 1987).

\bibitem{dietrich}
  D.~B\"odeker,
  {\it Moduli decay in the hot early Universe,}
  JCAP {06} (2006) 027
  [hep-ph/0605030].

\bibitem{landau9} 
  E.M.~Lifshitz and L.P.~Pitaevskii, 
  {\it Statistical Physics, Part 2}, \S88-89
  (Butterworth-Heinemann, Oxford, 1980).

\bibitem{sch}
  S.~Caron-Huot,
  {\it Hard thermal loops in the real-time formalism,}
  JHEP {04} (2009) 004
  [0710.5726].

\bibitem{guy}
  G.D.~Moore and K.A.~Sohrabi,
  {\it Kubo Formulae for Second-Order Hydrodynamic Coefficients,}
  Phys.\ Rev.\ Lett.\  {106} (2011) 122302
  [1007.5333].

\bibitem{mclerran}
  D.~B\"odeker, L.D.~McLerran and A.V.~Smilga,
  {\it Really computing nonperturbative real time correlation functions,}
  Phys.\ Rev.\ D {52} (1995) 4675
  [hep-th/9504123].

\bibitem{arnold}
  P.B.~Arnold,
  {\it Hot B violation, the lattice, and hard thermal loops,}
  Phys.\ Rev.\ D {55} (1997) 7781
  [hep-ph/9701393].

\bibitem{lat1}
  G.N.~Watson, 
 {\it Three triple integrals}, 
  Q.\ J.\ Math.\ 10 (1939) 266. 

\bibitem{lat2} 
 M.L.~Glasser and J.~Boersma,
 {\it Exact values for the cubic lattice Green functions}, 
  J.\ Phys.\ A:\ Math.\ Gen.\ 33 (2000) 5017.

\bibitem{jeon}
  S.~Jeon,
  {\it Hydrodynamic transport coefficients in
  relativistic scalar field theory,}
  Phys.\ Rev.\ D {52} (1995) 3591
  [hep-ph/9409250].

\bibitem{kovtun}
  P.~Kovtun, G.D.~Moore and P.~Romatschke,
  {\it The stickiness of sound: An absolute lower limit on viscosity
  and the breakdown of second order relativistic hydrodynamics,}
  Phys.\ Rev.\ D {84} (2011) 025006
  [1104.1586].

\bibitem{mr1}
  G.D.~Moore and K.~Rummukainen,
  {\it Electroweak bubble nucleation, nonperturbatively,}
  Phys.\ Rev.\ D {63} (2001) 045002
  [hep-ph/0009132].

\bibitem{classic}
 B.I.~Halperin, T.C.~Lubensky and S.-K.~Ma,
 {\it First-Order Phase Transitions in Superconductors and 
 Smectic-A Liquid Crystals},  
 Phys.\ Rev.\ Lett.\ 32 (1974) 292.

\bibitem{Huet}
  P.Y.~Huet, K.~Kajantie, R.G.~Leigh, B.H.~Liu and L.D.~McLerran,
  {\it Hydrodynamic stability analysis of burning bubbles
  in electroweak theory and in QCD,}
  Phys.\ Rev.\ D {48} (1993) 2477
  [hep-ph/9212224].

\bibitem{akamatsu}
  Y.~Akamatsu, A.~Mazeliauskas and D.~Teaney,
  {\it Bulk viscosity from hydrodynamic fluctuations with
  relativistic hydro-kinetic theory,}
  Phys.\ Rev.\ C {97} (2018) 024902
  [1708.05657].

\bibitem{meyer}
  L.~Giusti and H.B.~Meyer,
  {\it Implications of Poincar\'e symmetry for
  thermal field theories in finite-volume,}
  JHEP {01} (2013) 140
  [1211.6669].

\bibitem{kapusta}
  J.I.~Kapusta, B.~M\"uller and M.~Stephanov,
  {\it Relativistic Theory of Hydrodynamic Fluctuations with
  Applications to Heavy Ion Collisions,}
  Phys.\ Rev.\ C {85} (2012) 054906
  [1112.6405].

\bibitem{yaffe}
  P.~Kovtun and L.G.~Yaffe,
  {\it Hydrodynamic fluctuations, long time tails, and supersymmetry,}
  Phys.\ Rev.\ D {68} (2003) 025007
  [hep-th/0303010].

\end{thebibliography}
\end{document}